\documentclass[12pt]{article} 
 \usepackage{amsmath}
 \usepackage{amssymb}
 \usepackage{dsfont}
 \usepackage{amsthm}
   \font\twelvebm                       = cmmib10 at 12truept
   \font\tenbm                          = cmmib10 at 10truept
   \font\sevenbm                        = cmmib10 at 7truept
\tenbm \scriptscriptfont9              = \sevenbm

  at 10truept
 at 10truept  at 10truept  at 10truept 
at 10truept

\mathchardef \BGamma            = "0900 \mathchardef
\BDelta = "0901 \mathchardef \BTheta            = "0902
\mathchardef \BLambda           = "0903 \mathchardef \BXi
= "0904 \mathchardef \BPi               = "0905
\mathchardef \BSigma = "0906 \mathchardef \BUpsilon
= "0907 \mathchardef \BPhi = "0908 \mathchardef \BPsi
= "0909 \mathchardef \BOmega = "090A \mathchardef \Balpha
= "090B \mathchardef \Bbeta             = "090C
\mathchardef \Bgamma = "090D \mathchardef \Bdelta
= "090E \mathchardef \Bepsilon = "090F \mathchardef \Bzeta
= "0910 \mathchardef \Beta = "0911 \mathchardef \Btheta =
"0912 \mathchardef \Biota = "0913 \mathchardef \Bkappa
= "0914 \mathchardef \Blambda = "0915 \mathchardef \Bmu
= "0916 \mathchardef \Bnu = "0917 \mathchardef \Bxi
= "0918 \mathchardef \Bpi = "0919 \mathchardef \Brho
= "091A \mathchardef \Bsigma            = "091B
\mathchardef \Btau = "091C \mathchardef \Bupsilon
= "091D \mathchardef \Bphi = "091E \mathchardef \Bchi
= "091F \mathchardef \Bpsi = "0920 \mathchardef \Bomega
= "0921 \mathchardef \Bvarepsilon       = "0922
\mathchardef \Bvartheta         = "0923 \mathchardef
\Bvarpi = "0924 \mathchardef \Bvarrho = "0925 \mathchardef
\Bvarsigma = "0926 \mathchardef \Bvarphi           = "0927
\mathchardef \bA        = "0941 \mathchardef \bB        =
"0942 \mathchardef \bC        = "0943 \mathchardef \bD
= "0944 \mathchardef \bE        = "0945 \mathchardef \bF
= "0946 \mathchardef \bG        = "0947 \mathchardef \bH
= "0948 \mathchardef \bI        = "0949 \mathchardef \bJ
= "094A \mathchardef \bK        = "094B \mathchardef \bL
= "094C \mathchardef \bM        = "094D \mathchardef \bN
= "094E \mathchardef \bO        = "094F \mathchardef \bP
= "0950 \mathchardef \bQ        = "0951 \mathchardef \bR
= "0952 \mathchardef \bS        = "0953 \mathchardef \bT
= "0954 \mathchardef \bU        = "0955 \mathchardef \bV
= "0956 \mathchardef \bW        = "0957 \mathchardef \bX
= "0958 \mathchardef \bY        = "0959 \mathchardef \bZ
= "095A \mathchardef \ba        = "0961 \mathchardef \bb
= "0962 \mathchardef \bc        = "0963 \mathchardef \bd
= "0964
\mathchardef \bee       = "0965 
\mathchardef \bff       = "0966 \mathchardef \bg        =
"0967 \mathchardef \bh        = "0968
\mathchardef \bj        = "096A \mathchardef \bk        =
"096B \mathchardef \bl        = "096C \mathchardef \bm
= "096D \mathchardef \bn        = "096E \mathchardef \bo
= "096F \mathchardef \bp        = "0970 \mathchardef \bq
= "0971 \mathchardef \br        = "0972 \mathchardef \bs
= "0973 \mathchardef \bt        = "0974 \mathchardef \bu
= "0975 \mathchardef \bv        = "0976 \mathchardef \bw
= "0977 \mathchardef \bx        = "0978 \mathchardef \by
= "0979 \mathchardef \bz        = "097A

\font\tencb            = cmssbx10 scaled \magstep4
\font\eigcb = cmssbx10 scaled \magstep2 \textfont8
= \tencb \scriptfont8           = \eigcb \scriptscriptfont8
= \eigcb \mathchardef\bAs       = "1841
\def\Asem#1#2{\mathop{\vrule height10.5pt depth5.5pt width0pt\bAs}_{#1}^{#2}}
\def\asem#1#2{
          \ifmmode
         \ifinner
            \raise0.9pt\hbox{$\scriptstyle\bAs$}_{#1}^{#2}
         \else
            \Asem{#1}{#2}
         \fi
          \fi
          }

\newtheorem{theo}{\small\bf Theorem}[section]
\newenvironment{THEO}{\begin{theo} \rm}{\end{theo}}
\newtheorem{lem}{\small\bf Lemma}[section]
\newenvironment{LEM}{\begin{lem} \rm}{\end{lem}}
\newtheorem{prop}{\small\bf Proposition} 

\newtheorem{rem}{\small\bf Remark}[section]
\newenvironment{REM}{\begin{rem} \rm}{\end{rem}}
\newtheorem{defi}{\small\bf Definition}[section]

\newtheorem{cor}{\small\bf Corollary}[section]

\newtheorem{example}{\small\bf Example}[section]

\newenvironment{Proof}{Proof:}{}
\renewcommand{\Pr}{\mathds{P}}

\newcommand{\be}{\begin{equation}}
\newcommand{\ee}{\end{equation}}
\newcommand{\E}{\mathds{E}}
\newcommand{\Var}{\mbox{\rm \hspace*{.2ex}Var\hspace*{.2ex}}}
\newcommand{\Cov}{\mbox{\rm \hspace*{.2ex}Cov\hspace*{.2ex}}}

\newcommand{\Pol}{\mbox{\rm \hspace*{.2ex}Pol\hspace*{.2ex}}}
\newcommand{\sign}{\mbox{\rm{sign}}}

\newcommand{\R}{\mathds{R}}

\newcommand{\dist}{\stackrel{\mbox{\footnotesize d}}{=}}

 \newcommand{\xwrosA}{$\begin{array}{c}\vspace{-2ex}\\}
 \newcommand{\xwrosB}{\\ \vspace{-2ex}\end{array}$}

 \title{\bf A Simple Method for Obtaining the Maximal Correlation
 Coefficient and Related
 Characterizations\footnote{Work partially supported by the University of
 Athens Research Grant 70/4/5637.}}

 \author{\large
 Nickos Papadatos\footnote{\mbox{{\it Corresponding author.}
 e-mail:\ {\tt npapadat@math.uoa.gr} \ url:\ {\tt http://users.uoa.gr/$\sim$npapadat/}}} \ \ and \ \
 Tatiana
 Xifara\footnote{e-mail:\ {\tt  t.xifara@lancaster.ac.uk} \
 url:\ {\tt http://www.maths.lancs.ac.uk/$\sim$xifara/}}}

 \date{\small
 Section of Statistics and O.R.,
 Department of Mathematics,
 University of Athens,
 \\ Panepistemiopolis, 157 84 Athens, Greece
 \\
 and
 \\
 Department of Mathematics and Statistics, Lancaster University, UK}

\begin{document}

 \maketitle


 \vspace{-1em}

 \thispagestyle{empty}

 \begin{abstract}
 \noindent
 We provide a method that
 enables the simple calculation of the maximal correlation
 coefficient of a bivariate distribution, under suitable
 conditions. In particular, the method readily applies to known
 results on order statistics and records. As an application we
 provide a new characterization of the exponential distribution:
 Under a splitting model on independent identically distributed observations,
 it is the (unique, up to a location-scale transformation) parent distribution
 that maximizes the correlation coefficient between the records among two
 different branches of the splitting sequence.
 \end{abstract}
 {\footnotesize {\it MSC}:  Primary 62H20, 62G30; Secondary 62E10, 60E15.

 }

 \vspace*{-1em}

 \section{Introduction}
 \label{sec.intro}
 As is well-known, the {\it Pearson correlation
 coefficient}
 of the random variables (r.v.'s) $X$ and $Y$ is defined
 as
 \[
 \rho(X,Y)=
 \frac{\Cov(X,Y)}{\sqrt{\Var(X)}\sqrt{\Var(Y)}},
 \]
 provided that $0<\Var(X)<\infty$ and $0<\Var(Y)<\infty$.
 It assumes values in the interval $[-1,1]$ and it is a
 measure of {\it linear dependence} of $X$ and $Y$.
 Although
 $\rho(X,Y)=0$ for independent $X$ and $Y$, the converse is not
 true. Gebelein (1941) introduced the {\it maximal correlation
 coefficient},
 \[
 R(X,Y)=\sup \rho\big( g_{1}(X),g_{2}(Y)\big),
 \]
 where the supremum is taken over
 all Borel functions
 $g_{1}:\R\to\R$ and  $g_{2}:\R\to\R$  with $0<\Var g_{1}(X)<\infty$ and
 $0<\Var g_{2}(Y)<\infty$. In contrast to
 $\rho(X,Y)$, $R(X,Y)$ is defined whenever
 both $X$ and $Y$ are non-degenerate, assumes values
 in the interval $[0,1]$ and vanishes
 if and only if $X$ and $Y$ are independent.
 The maximal correlation coefficient plays a fundamental role in
 various areas of statistics; e.g.,
 it is useful in obtaining optimal transformations
 for regression, Breiman and Friedman (1985), and it has
 applications in the convergence theory of Gibbs sampling
 algorithms, Liu et al.\ (1994).

 However, despite its usefulness, it is often difficult to calculate
 the maximal correlation coefficient in an explicit form, except
 in some rare cases. A well-known exception is the
 result of Gebelein (1941)
 and Lancaster (1957)
 who show that
 if $(X,Y)$ is bivariate normal then
 \be
 \label{normal}
 R(X,Y)=|\rho(X,Y)|.
 \ee
 Another exception is provided by the surprising result of Dembo et al.\ (2001),
 and its subsequent extensions given by Bryc et al.\ (2005) and Yu
 (2008).
 In its general form the result states that for
 any independent identically distributed (i.i.d.) non-degenerate
 r.v.'s
 $X_1,\ldots,X_n$,
 \[
 R(X_1+\cdots+X_m,X_{k+1}+\cdots+X_n)=\frac{m-k}{\sqrt{m(n-k)}},
 \ \ 1\leq k+1\leq m\leq n.
 \]
 Finally, we mention an important result of Sz\'{e}kely and M\'{o}ri
 (1985), who showed, using Jacobi polynomials, that if $(X,Y)$ follows a bivariate
 density of the form
 \be
 \label{eq.2}
 f(x,y)=\frac{\Gamma(\alpha+\beta+\gamma)}
 {\Gamma(\alpha)\Gamma(\beta)\Gamma(\gamma)}x^{\alpha-1}(y-x)^{\beta-1}(1-y)^{\gamma-1},
 \ \ 0<x<y<1,
 \ee
 where the parameters $\alpha$, $\beta$, $\gamma$ are
 positive, then
 \be
 \label{eq.3}
 R(X,Y)=\rho(X,Y)=\sqrt{\frac{\alpha\gamma}{(\beta+\alpha)(\beta+\gamma)}}.
 \ee
 Observe that for any integers $1\leq i<j\leq n$,
 the density of the pair of order statistics
 $(U_{i:n},U_{j:n})$,
 based on $n$ i.i.d.\ observations from the standard
 uniform distribution, is of the form (\ref{eq.2})
 with $\alpha=i$, $\beta=j-i$, $\gamma=n+1-j$. Actually,
 (\ref{eq.3}) extends Terrell's (1983) characterization of
 rectangular distributions through maximal correlation
 of an ordered pair.

 In this article we provide a unified method for obtaining the
 maximal correlation coefficient when the bivariate distribution has a
 particular diagonal structure  -- see next section.
 The method is very simple
 (e.g., it readily applies to verify (\ref{normal}) and (\ref{eq.3}))
 and it does not require knowledge of particular sets of orthogonal
 polynomials.
 Section \ref{sec.known} presents some notable examples
 of known characterizations
 of specific distributions through maximal correlation of ordered data
 and records.
 We consider a splitting model
 based on i.i.d.\ observations in Section \ref{sec.last}.
 Applying our method it is shown
 that the records among two different branches of the splitting
 sequence are maximally correlated if and only if the population
 distribution is exponential (up to location-scale transformations) --
 this extends Nevzorov's (1992) characterization.

 \section{The maximal correlation coefficient
 of bivariate distributions having diagonal structure}
 \label{sec.newmethod}
 Let $(X,Y)$ be an arbitrary
 random vector with distribution function $F(x,y)$
 and
 assume that both $X$
 and $Y$ are non-degenerate.
 We say that $F$, similarly the vector $(X,Y)$,
 has diagonal structure if the following
 three conditions are
 satisfied.
 \smallskip

 \noindent
 {\small\bf A1.} We assume that both $X$ and $Y$ have all their moments
 finite:
 \be
 \label{a1}
 \E |X|^n<\infty \ \mbox{and} \ \E|Y|^n<\infty \
 \mbox{ for } n=1,2,\ldots \ .
 \ee

 It is known that, under (\ref{a1}), there exists a
 (unique) orthonormal polynomial system (OPS)
 $\{\phi_n(x)=p_n x^n +\Pol_{n-1}(x), \ p_n>0, n=0,1,\ldots\}$,
 corresponding to $X$, and a (unique) OPS
 $\{\psi_n(y)=q_n y^n +\Pol_{n-1}(y), \ q_n>0, n=0,1,\ldots\}$,
 corresponding to $Y$. Here $\phi_0(x)\equiv \psi_0(y)\equiv 1$
 and $\Pol_{k}(t)$ denotes an arbitrary polynomial in $t$ of
 degree less than or equal to $k$, which may change from line
 to line. The orthonormality of the above OPS's means, as usual, that
 \[
 \E[\phi_n(X)\phi_k(X)]=\E[\psi_n(Y)\psi_k(Y)]=\delta_{kn}, \ \
 k,n=0,1,\ldots,
 \]
 where $\delta_{kn}$ is Kronecker's delta.

 \begin{REM}
 \label{rem.new}
 For a random variable $X$ we denote by $\nu_X+1$
 the cardinality of its (minimal closed) support, $S(X)$,
 unifying the cases where $\nu_X<\infty$ and $\nu_X=\infty$.
 This convention is necessary because the OPS,
 corresponding to a non-degenerate
 r.v.\ $X$, reduces
 to the finite set $\{\phi_n(x)\}_{n=0}^{\nu_X}$
 if (and only if) its support is concentrated
 on a finite subset
 of $\R$, with $\nu_X+1\geq 2$ points.
 This singular case,
 however, appears in some interesting situations
 -- e.g., see Section \ref{sec.known}, below, regarding
 the finite population case.
 In order to fix this problem
 (and give a unified presentation of the results)
 we shall proceed as follows.
 In any case where the support of $X$ is
 of form
 $\{x_0,x_1,\ldots,x_{\nu_X}\}$,
 we shall enlarge the finite set of orthonormal polynomials to
 $\{\phi_n\}_{n=0}^{\infty}$,
 keeping $\{\phi_n\}_{n=0}^{\nu_X}$ as above,
 and defining
 \[
 \mbox{
 $\phi_n(x):=x^{n-\nu_X-1}(x-x_0)(x-x_1)\cdots
 (x-x_{\nu_X})$,
 \ \ $n>\nu_X$.
 }
 \]
 Each $\phi_n$ in the enlarged set is of degree $n$ and has principal
 coefficient $p_n>0$. However, since for $n>\nu_X$, $\phi_n(X)=0$ w.p.\ 1,
 the orthonormality assumption has now been relaxed to
 \[
 \E[\phi_n(X)\phi_k(X)]=\delta_{kn} {\bf1}_{\{n\leq \nu_X\}},
 \ \
 k,n=0,1,\ldots,
 \]
 where $\bf1$ stands for the indicator function.
 The same conventions will be applied to the OPS of
 $Y$, by setting
 $\psi_n(y):=y^{n-\nu_Y-1}(y-y_0)\cdots(y-y_{\nu_Y})$,
 whenever $n> \nu_Y$ and $S(Y)=\{y_0,\ldots,y_{\nu_Y}\}$ is finite.
 \end{REM}

 \medskip

 \noindent
 {\small\bf A2.} We assume that the
 OPS $\{\phi_n(x)\}_{n=0}^{\infty}$  is
 complete in $L^2(X)$, the Hilbert space of all
 Borel functions $g:\R\to\R$ with $\Var g(X)<\infty$.
 Clearly, the enlarged OPS of Remark \ref{rem.new}
 is complete if and only if the ordinary OPS is,
 noting that two functions $g_1$, $g_2$ are
 considered as ``equal" if $\Pr[g_1(X)=g_2(X)]=1$. Similarly,
 we assume that the system $\{\psi_n(y)\}_{n=0}^{\infty}$
 is complete in $L^2(Y)$.
 \medskip

 \noindent
 {\small\bf A3.} We assume that the random vector $(X,Y)$
 has the {\it polynomial regression property}. That is,
 \begin{eqnarray*}
 \E(X^{n}|Y) &=& A_{n}Y^{n}+\Pol_{n-1}(Y), \ \  n=1,2,\ldots,
 \\
 \E(Y^{n}|X) &=&  B_{n}X^{n}+\Pol_{n-1}(X), \ \ n=1,2,\ldots,
 \end{eqnarray*}
 where $A_{n},B_{n}\in \R$.
 \medskip

 The assumptions {\small\bf A1} and {\small\bf A2} are not
 very restrictive.
 For example, they are satisfied whenever both $X$
 and $Y$ have finite moment generating functions in a
 neighborhood of $0$; see, for example, Koudou (1998) and
 Afendras et al.\ (2011). However, this is not the case for
 assumption {\small\bf A3}, since it applies
 to very particular distributions, as the following
 lemma shows.

 \begin{LEM}
 \label{lem.1}
 Using the above notation and assuming
 {\small\bf A1}--{\small\bf A3} we have
 that for all $n,k\in\{1,2,\ldots\}$,
 \be
 \label{eq.5}
 \E[\phi_n(X)\psi_k(Y)]=\delta_{nk} \rho_n,
 \ee
 where $\delta_{nk}$ is Kronecker's delta and
 $\rho_n:=\E[\phi_n(X)\psi_n(Y)]\in[-1,1]$.
 \end{LEM}

 \noindent
 \begin{Proof} Set $\nu=\min\{\nu_X,\nu_Y\}\in\{1,2,\ldots\}\cup\{\infty\}$
 (for the definition of $\nu_X, \nu_Y$ see Remark \ref{rem.new}).
 If $n\leq \nu$
 then $\phi_n(X)$ and $\psi_n(Y)$
 are standardized r.v.'s, and we have
 $\rho_n=\rho(\phi_n(X),\psi_n(Y))$. Therefore,
 $\rho_n\in[-1,1]$ in this case. If at least one
 of $\nu_X$, $\nu_Y$ is finite then for every $n>\nu$,
 $\phi_n(X)\psi_n(Y)=0$ w.p.\ 1 , so that
 $\rho_n=0$ for $n>\nu$. Thus, $\rho_n\in[-1,1]$ for all $n$.
 Now, if $1\leq k<n$ then {\small\bf A3} yields
 \[
 \E[\phi_n(X)\psi_k(Y)]=\E\{\phi_n(X)\E[\psi_k(Y)|X]\}=\E[\phi_n(X)\Pol_k(X)]=0,
 \]
 because $\phi_n$ is orthogonal to any polynomial of degree at
 most $n-1$. Similar arguments apply to the case
 $1\leq n<k$, and the proof is complete.
 \medskip
 $\Box$
 \end{Proof}

 The bivariate distributions satisfying (\ref{eq.5})
 are sometimes called
 {\it Lancaster distributions} and the correlations
 $\rho_n$ form a {\it Lancaster sequence} with respect to
 $X$ and $Y$; see Lancaster (1969), Koudou (1996, 1998).
 Therefore, by Lemma \ref{lem.1} we see that assumption
 {\small\bf A3}
 forces a distribution to be a Lancaster one.
 Under certain conditions, the density of a Lancaster distribution,
 if it exists, has the formal representation (diagonal structure)
 \[
 f(x,y)=f_X(x) f_Y(y) \Big(1+\sum_{n=1}^\infty \rho_n
 \phi_n(x)\psi_n(y)\Big),
 \]
 where $f_X$ and $f_Y$ are the marginal densities of $X$ and
 $Y$.

 If the assumptions {\small\bf A1}--{\small\bf A3} are satisfied
 then we can calculate
 each $\rho_n$, and this calculation does not require
 any knowledge of the polynomial systems $\{\phi_n(x)\}_{n=0}^{\infty}$
 and $\{\psi_n(y)\}_{n=0}^{\infty}$. Indeed, we have the following

 \begin{LEM}
 \label{lem.2}
 Let $\nu=\min\{\nu_X,\nu_Y\}$ (see Remark \ref{rem.new}).
 Using the above notation and assuming {\small\bf A1}--{\small\bf A3} we have
 that for all $n\in\{1,2,\ldots\}$,
 \be
 \label{eq.6}
 A_nB_n {\bf1}_{\{n\leq \nu\}}\geq 0, \ \
 \rho_n=\sign(A_n)\sqrt{A_nB_n{\bf1}_{\{n\leq \nu\}}} \ \ \mbox{and} \ \
 |\rho_n|=\sqrt{A_nB_n{\bf1}_{\{n\leq \nu\}}}.
 \ee
 \end{LEM}

 \noindent
 \begin{Proof}
 Since $\phi_{n}(X)=p_{n}X^{n}+\Pol_{n-1}(X)$ and
 $ \psi_{n}(Y)=q_{n}Y^{n}+\Pol_{n-1}(Y)$ we have
 \begin{eqnarray*}
 \rho_{n}
 \hspace{-1ex}& = &\hspace{-1ex}
 \E\{\psi_n(Y)\E(\phi_{n}(X)|Y)\}
 =
 \E\{\psi_{n}(Y)[p_{n}\E(X^{n}|Y)+\Pol_{n-1}(Y)]\}
 \vspace{.5ex}\\
 \hspace{-1ex}& = &\hspace{-1ex}
 p_{n}\E[\psi_{n}(Y)\E(X^{n}|Y)]+0=p_{n}\E\{\psi_{n}(Y)[A_{n}Y^{n}+\Pol_{n-1}(Y)]\}
 \vspace{.5ex}\\
 \hspace{-1ex}& = &\hspace{-1ex}
 p_{n}A_{n}\E[\psi_{n}(Y)Y^{n}]+0=p_{n}A_{n}
 \E\{\psi_{n}(Y)q_n^{-1}[\psi_n(Y)-\Pol_{n-1}(Y)]\}
 \vspace{.5ex}\\
 \hspace{-1ex}& = &\hspace{-1ex}
 \mbox{$\frac{p_{n}A_{n}}{q_{n}}\E[\psi_{n}^{2}(Y)]-0
 =\frac{p_{n}A_{n}}{q_{n}}{\bf1}_{\{n\leq \nu_Y\}}$}.
 \end{eqnarray*}
 This shows that $\rho_n$ and $A_n{\bf1}_{\{n\leq \nu_Y\}}$
 have the same sign (in particular, $\rho_n=0$ for $n>\nu_Y$).
 Using the same arguments (conditioning on $X$) it
 follows that $\rho_n=\frac{q_n B_n}{p_n}{\bf1}_{\{n\leq \nu_X\}}$;
 thus, $\rho_n=0$ for $n>\nu_X$. Therefore, if $\nu$ is finite then
 $\rho_n=0$ for all $n>\nu$.
 Finally, $\rho_n^2=A_n B_n{\bf1}_{\{n\leq \nu_X\}}{\bf1}_{\{n\leq \nu_Y\}}
 =A_nB_n{\bf1}_{\{n\leq \nu\}}$, and the proof is complete.
 \medskip
 $\Box$
 \end{Proof}

 We are now in a position to state and prove our main result.
 \begin{THEO}
 \label{th.1}
 If the assumptions {\small\bf A1}--{\small\bf A3} are satisfied
 and $\nu=\min\{\nu_X,\nu_Y\}$ then
 \be
 \label{eq.12}
 R(X,Y)=\sup_{n\geq 1} |\rho_n|=\sup_{n\geq 1}\sqrt{A_nB_n{\bf1}_{\{n\leq \nu\}}}.
 \ee
 Moreover, if $|\rho_n|<|\rho_{n_0}|$ for all
 $n\geq 1$, $n\neq n_0$,
 then for any $g_1\in L^2(X)$ with $\Var g_1(X)>0$ and for
 any $g_2\in L^2(Y)$ with $\Var g_2(Y)>0$ we
 have the inequality
 \[
 \rho(g_1(X),g_2(Y))\leq |\rho_{n_0}|=\sqrt{A_{n_0}B_{n_0}},
 \]
 with equality if and only if
 $g_1(x)=a_0+a_1 \phi_{n_0}(x)$
 and
 $g_2(y)=b_0+b_1 \psi_{n_0}(y)$
 for some constants
 $a_0,b_0,a_1,b_1\in\R$ with
 $a_1b_1\sign(A_{n_0})>0$.
 \end{THEO}

 \noindent
 \begin{Proof}
 Let $g_1\in L^2(X)$ and denote by $F_X$ the marginal distribution function
 of $X$. By the completeness of $\{\phi_n\}_{n=0}^{\infty}$ it follows
 that $g_1$ admits the representation
 \[
 g_1(x)=\sum_{n=0}^{\infty} \alpha_n \phi_n(x), \ \
 \mbox{where} \ \alpha_n=\E[g_1(X)\phi_n(X)]=\int_{\R} g_1(x)
 \phi_n(x) dF_X(x).
 \]
 Clearly, if $\nu_X$ is
 finite and $n>\nu_X$ then $\alpha_n=0$, because
 $\Pr(\phi_n(X)=0)=1$;
 see Remark \ref{rem.new}.
 The constants
 $\{\alpha_n\}_{n=0}^{\infty}$
 are the Fourier coefficients of $g_1$ with respect to the OPS $\{\phi_n\}_{n=0}^{\infty}$,
 and the series converges in the $L^2(X)$-sense, i.e.,
 \be
 \label{eq.7}
 \lim_{N} \E
 \Big[g_1(X)-\sum_{n=0}^N \alpha_n \phi_n(X)\Big]^2=0.
 \ee
 In particular, $\alpha_0=\E[g_1(X)]$, and the above limit is
 usually written as Parseval's identity,
 \[
 \Var g_1(X)=\sum_{n=1}^{\infty} \alpha_n^2
 \]
 (equivalently, $\Var g_1(X)=\sum_{n=1}^{\nu_X}\alpha_n^2$ if $\nu_X<\infty$),
 since it is easily verified that
 \[
 \E
 \Big[g_1(X)-\sum_{n=0}^N \alpha_n \phi_n(X)\Big]^2
 = \Var g_1(X)-\sum_{n=1}^{N} \alpha_n^2.
 \]
 Therefore, the assumption $\Var g_1(X)>0$ implies that $\alpha_n\neq 0$ for at
 least one $n\geq 1$. Similarly, for any
 $g_2\in L^2(Y)$ we have
 \[
 \Var g_2(Y)=\sum_{n=1}^{\infty} \beta_n^2, \ \
 \mbox{where} \ \beta_n=\E[g_2(Y)\psi_n(Y)]=\int_{\R} g_2(y)
 \psi_n(y) dF_Y(y),
 \]
 where $F_Y$ is the marginal distribution of $Y$, $\{\beta_n\}_{n=0}^{\infty}$
 are the Fourier coefficients of $g_2$ with respect to the OPS $\{\psi_n\}_{n=0}^{\infty}$
 ($\beta_n=0$ if $\nu_Y<\infty$ and $n>\nu_Y$) and, as for $X$,
 \be
 \label{eq.8}
 \lim_{N}
 \E
 \Big[g_2(Y)-\sum_{n=0}^N \beta_n \psi_n(Y)\Big]^2
 =\Var g_2(Y)-\lim_{N}\sum_{n=1}^N\beta_n^2 = 0.
 \ee
 Using the above we can show that
 \be
 \label{eq.9}
 \E[g_1(X)\psi_n(Y)]=\alpha_n \rho_n
 \ \ \
 \mbox{and}
 \ \ \
 \E[g_2(Y)\phi_n(X)]=\beta_n
 \rho_n, \ \ \ n=1,2,\ldots \  .
 \ee
 Indeed, for any $N\geq n$ we have
 \[
 \E[g_1(X)\psi_n(Y)]=
 \E\Big\{\Big[g_1(X)-\sum_{k=0}^N \alpha_k \phi_k(X)\Big]\psi_n(Y)\Big\}+
 \sum_{k=0}^N \alpha_k \E[\phi_k(X)\psi_n(Y)].
 \]
 Now $N\geq n$, $\phi_0(x)\equiv 1$, $\E[\psi_n(Y)]=0$,
 $\E[\psi_n^2(Y)]={\bf1}_{\{n\leq \nu_Y\}}$
 and
 $\E[\phi_k(X)\psi_n(Y)]=\delta_{kn}\rho_n$ for $k\geq 1$.
 Thus, in view of (\ref{eq.7}) and by the Cauchy-Schwarz inequality,
 \begin{eqnarray*}
 &&
 \hspace{-7ex}
 0\leq\Big(\E[g_1(X)\psi_n(Y)]-\alpha_n\rho_n\Big)^2=
 \Big(\E\Big\{\Big[g_1(X)-\sum_{k=0}^N \alpha_k
 \phi_k(X)\Big]\psi_n(Y)\Big\}\Big)^2 \\
 &&
 \leq \E
 \Big[g_1(X)-\sum_{k=0}^N \alpha_k
 \phi_k(X)\Big]^2\E[\psi_n^2(Y)]
 \to 0, \ \ \mbox{as} \ N\to\infty.
 \end{eqnarray*}
 Therefore, since
 $\left(\E[g_1(X)\psi_n(Y)]-\alpha_n\rho_n\right)^2$ does not
 depend on $N$, we conclude the first identity in
 (\ref{eq.9}). The remainder of (\ref{eq.9}) follows
 similarly.
 From (\ref{eq.9}) we obtain
 \begin{eqnarray*}
 \E
 \Big[\Big(g_1(X)-\sum_{n=0}^N \alpha_n\phi_n(X)\Big)
 \Big(g_2(Y)-\sum_{n=0}^N \beta_n\psi_n(Y)\Big)\Big] \\
 =\Cov[g_1(X),g_2(Y)]-\sum_{n=1}^N \rho_n\alpha_n\beta_n.
 \end{eqnarray*}
 Thus, squaring the above identity and applying the
 Cauchy-Schwarz inequality to the resulting squared expectation we
 conclude, in view of (\ref{eq.7}) and (\ref{eq.8}), that
 \be
 \label{eq.10}
 \Cov[g_1(X),g_2(Y)]=
 \sum_{n=1}^\infty
 \rho_n\alpha_n\beta_n
 =
 \sum_{n=1}^{\nu}
 \rho_n\alpha_n\beta_n.
 \ee
 [Recall that $\nu=\min\{\nu_X,\nu_Y\}$; for the definition of
 $\nu_X,\nu_Y$ see Remark
 \ref{rem.new}.]\
 Therefore, combining the above we obtain the expression
 \be
 \label{eq.11}
 \rho(g_{1}(X),g_{2}(Y))=
 \frac{
 \sum_{n=1}^{\infty}\rho_n\alpha_{n}\beta_{n}}
 {
 \sqrt{\sum_{n=1}^{\infty}\alpha_{n}^{2}}
 \sqrt{\sum_{n=1}^{\infty}\beta_{n}^{2}}}
 =
 \frac{
 \sum_{n=1}^{\nu}\rho_n\alpha_{n}\beta_{n}}
 {
 \sqrt{\sum_{n=1}^{\nu_X}\alpha_{n}^{2}}
 \sqrt{\sum_{n=1}^{\nu_Y}\beta_{n}^{2}}}.
 \ee
 Observe that, in view of (\ref{eq.10}),
 \begin{eqnarray*}
 \Big(\Cov[g_1(X),g_2(Y)]\Big)^2
 \hspace{-1ex}& = &\hspace{-1ex}
 \Big|\sum_{n=1}^\infty
 \rho_n\alpha_n\beta_n\Big|^2
 \leq
 \Big(\sum_{n=1}^\infty
 |\rho_n||\alpha_n||\beta_n|\Big)^2
 \\
 \hspace{-1ex}& = &\hspace{-1ex}
 \Big(\sum_{n=1}^\infty
 (\sqrt{|\rho_n|}|\alpha_n|)(\sqrt{|\rho_n|}|\beta_n|)\Big)^2
 \\
 \hspace{-1ex}& \leq &\hspace{-1ex}
 \Big(\sum_{n=1}^\infty |\rho_n|\alpha_n^2\Big)
 \Big(\sum_{n=1}^\infty |\rho_n|\beta_n^2\Big)
 \\
 \hspace{-1ex}& \leq &\hspace{-1ex}
 \Big(\big(\sup_{n\geq 1}|\rho_n|\big)\sum_{n=1}^\infty \alpha_n^2\Big)
 \Big(\Big(\sup_{n\geq 1}|\rho_n|\Big)\sum_{n=1}^\infty \beta_n^2\Big)
 \\
 \hspace{-1ex}& = &\hspace{-1ex}
 \big(\sup_{n\geq 1}\rho_n^2\big)\Big(\sum_{n=1}^\infty \alpha_n^2\Big)
 \Big(\sum_{n=1}^\infty \beta_n^2\Big).
 \end{eqnarray*}
 The above inequality, combined with (\ref{eq.11}), shows
 that
 \[
 R(X,Y)\leq \sup_{n\geq 1}|\rho_n|=R, \ \mbox{say}.
 \]
 On the other hand, for any $\epsilon>0$ we can find an
 index $n_0$ (with $n_0\leq \nu$ if $\nu$ is finite)
 such that $|\rho_{n_0}|>R-\epsilon$, and thus,
 $|\rho(\phi_{n_0}(X),\psi_{n_0}(Y))|=|\rho_{n_0}|>R-\epsilon$.
 Therefore,
 \begin{eqnarray*}
 R(X,Y)
 \hspace{-1ex}&=&\hspace{-1ex}
 \sup \rho(g_1(X),g_2(Y))
 \\
  \hspace{-1ex}&\geq&\hspace{-1ex}
 \max\{\rho(\phi_{n_0}(X),\psi_{n_0}(Y)),
 \rho(-\phi_{n_0}(X),\psi_{n_0}(Y))\}
 \\
 \hspace{-1ex}&=&\hspace{-1ex}
 \max\{\rho_{n_0},-\rho_{n_0}\}=|\rho_{n_0}|>R-\epsilon.
 \end{eqnarray*}
 Since the inequality $R(X,Y)>R-\epsilon$ holds for all
 $\epsilon>0$ it follows that $R(X,Y)\geq R$,
 and thus, $R(X,Y)=R$. It is clear that
 if the sequence $\{|\rho_n|\}_{n=1}^{\infty}$ has a unique
 maximum, say $|\rho_{n_0}|>0$, then it is necessary that
 $n_0\leq \nu$ if $\nu$ is finite. Working as above, it is
 easily shown that
 \[
 \Big(\Cov[g_1(X),g_2(Y)]\Big)^2
 \leq
 \rho_{n_0}^2\Big(\sum_{n=1}^\infty \alpha_n^2\Big)
 \Big(\sum_{n=1}^\infty \beta_n^2\Big)
 =\rho_{n_0}^2 \Var g_1(X) \Var g_2(Y),
 \]
 with equality if and only if $\alpha_n=\beta_n=0$ for all
 $n\geq 1$, $n\neq n_0$.
 Combining this with the fact that $\rho_{n_0} \ (=\rho(\phi_{n_0}(X),\psi_{n_0}(Y)))$
 has the sign of $A_{n_0}$, completes the proof.
 \medskip
 $\Box$
 \end{Proof}

 \section{Examples providing known characterizations via maximal correlation}
 \label{sec.known}
 The following known results are immediate applications of Theorem
 \ref{th.1}.
 \smallskip

 \noindent
 {\small\bf The bivariate normal case.}
 Assumptions {\small\bf A1}--{\small\bf A3} are easily checked
 for a bivariate
 normal. Indeed, if $(X,Y)$ is bivariate normal with
 $\E(X)=\mu_1$, $\E(Y)=\mu_2$,
 $\Var(X)=\sigma_1^2>0$, $\Var(Y)=\sigma_2^2>0$ and
 $\rho(X,Y)=\rho\in[-1,1]$
 then it is well-known that
 $(X|Y=y)\sim {\cal N}(\mu_1+\rho\frac{\sigma_1}{\sigma_2}(y-\mu_2),
 (1-\rho^2)\sigma_1^2)$. It follows that
 \[
 (X|Y=y) \dist
 \mu_1+\rho\frac{\sigma_1}{\sigma_2}(y-\mu_2)
 +\sigma_1\sqrt{1-\rho^2} Z,
 \]
 where $Z\sim {\cal N}(0,1)$ and $\dist$ denotes equality in
 distribution. Therefore,
 \[
 \E[X^n|Y=y]=\E\Big[\mu_1+\rho\frac{\sigma_1}{\sigma_2}(y-\mu_2)
 +\sigma_1\sqrt{1-\rho^2} Z\Big]^n=\rho^n
 \frac{\sigma_1^n}{\sigma_2^n} y^n + \Pol_{n-1}(y).
 \]
 That is,
 \[
 \E(X^n|Y)=A_n Y^n +\Pol_{n-1}(Y), \ \mbox{where} \  A_n=\rho^n
 \frac{\sigma_1^n}{\sigma_2^n}, \  n=1,2,\ldots \  .
 \]
 Similarly, $\E(Y^n|X)=B_n X^n +\Pol_{n-1}(X)$, where
 $B_n=\rho^n\frac{\sigma_2^n}{\sigma_1^n}$ for all $n\geq
 1$.
 Thus, {\small\bf A3} is satisfied, while {\small\bf A1} and
 {\small\bf A2} are well-known for the normal law
 (the moment generating function is finite). Since
 $\nu=\infty$, it
 follows from (\ref{eq.6})
 that $|\rho_n|=\sqrt{A_nB_n}=|\rho|^n$,
 $\rho_n=\sign(\rho^n)|\rho|^n=\rho^n$,
 and, by (\ref{eq.12}),
 $R(X,Y)=\sup_{n\geq 1}|\rho_n|=\max_{n\geq 1}|\rho|^n=|\rho|$.
 Moreover, in the particular case where $0<|\rho|<1$, the equality in
 \[
 |\rho(g_1(X),g_2(Y))|\leq |\rho|
 \]
 is attained if and only if both $g_1$, $g_2$ are linear.
 It is worth noting that (\ref{eq.10})
 takes the simple form (holding for any $\rho\in[-1,1]$)
 \be
 \label{eq.stein}
 \Cov[g_1(X),g_2(Y)]=\sum_{n=1}^\infty
 \frac{\rho^n\sigma_1^n\sigma_2^n}{n!}\E[g_1^{(n)}(X)]\E[g_2^{(n)}(Y)],
 \ee
 provided that $g_1,g_2\in C^{\infty}$, $g_1(X)\in L^2(X)$, $g_2(Y)\in L^2(Y)$,
 and that $\E|g_1^{(n)}(X)|<\infty$
 and $\E|g_2^{(n)}(Y)|<\infty$ for all $n$, where
 $g_i^{(n)}$ denotes the $n$-th derivative of $g_i$,
 $i=1,2$. Of course, one can apply (\ref{eq.stein}) to the
 case $X=Y$. Then, $\mu_1=\mu_2=\mu$, say, $\rho=1$,
 and $\sigma_1=\sigma_2=\sigma$, say, and (\ref{eq.stein}) yields
 the generalized Stein identity for the ${\cal N}(\mu,\sigma^2)$
 distribution (see Afendras et al.\ (2011)):
 \[
 \Cov[g_1(X),g_2(X)]=\sum_{n=1}^\infty
 \frac{(\sigma^2)^n}{n!}\E[g_1^{(n)}(X)]\E[g_2^{(n)}(X)].
 \]

 \noindent
 {\small\bf Characterization of rectangular distributions via maximal correlation
 of order statistics.} Terrell (1983), using Legendre
 polynomials, proved that if $X_{1:2}\leq X_{2:2}$ are
 the order statistics of two i.i.d.\ observations from a distribution with
 finite variance then
 \[
 \rho(X_{1:2},X_{2:2})\leq \frac{1}{2},
 \]
 and the equality characterizes the rectangular (uniform
 over some non-degenerate interval) distributions.
 However, Theorem \ref{th.1} applies immediately here. Indeed, if
 ${\cal U}(a,b)$ denotes the uniform distribution over $(a,b)$
 and  $U_1,U_2\sim{\cal U}(0,1)$ then it is obvious that
 the order statistics of $U_1$, $U_2$, $U_{1:2}\leq U_{2:2}$, satisfy the
 \medskip
 following:

 \noindent
 \begin{tabular}{r c l}
 $U_{1:2}|U_{2:2}\sim {\cal U}(0,U_{2:2}) $ & $\Rightarrow$ & $
 \E\big(U_{1:2}^{n}|U_{2:2}\big)
 =
 \int_{0}^{U_{2:2}}t^{n}\frac{1}{U_{2:2}}dt=\frac{1}{n+1}U_{2:2}^{n},$
 \vspace{.7em}
 \\
 $U_{2:2}|U_{1:2}\sim {\cal U}(U_{1:2},1)$ & $\Rightarrow$ & $
 \E\big(U_{2:2}^{n}|U_{1:2}\big)
 =
 \int_{U_{1:2}}^{1}t^{n}\frac{1}{1-U_{1:2}}dt$\vspace{0.1cm}\\
 & & $\hspace{12.3ex}=
 \frac{1}{n+1}(1+U_{1:2}+\cdots+U_{1:2}^{n}).$
 \end{tabular}
 \medskip
 \\
 Thus, $A_{n}=B_{n}=\frac{1}{n+1}$ and $|\rho_n|=\frac{1}{n+1}$.
 Therefore, $\max_{n\geq 1}|\rho_n|=|\rho_1|=\frac{1}{2}$.
 It follows from Theorem \ref{th.1} that
 $\rho(g(U_{1:2}),g(U_{2:2}))\leq \frac{1}{2}$,
 with equality if and only if $g$ is linear. Since for
 order statistics $X_{1:2}\leq X_{2:2}$ from an arbitrary
 distribution $F$
 \[
 (X_{1:2}, X_{2:2})\dist (g(U_{1:2}),
 g(U_{2:2})), \ \mbox{where} \
 g(u)=\inf\{x:F(x)\geq u\}, \ 0<u<1,
 \]
 (the above $g$ is usually denoted as $F^{-1}$),
 Terrell's result follows. The above argument
 can easily be extended to provide the characterization of
 Sz\'{e}kely and M\'{o}ri (1985), concerning the order statistics
 $X_{1:n}\leq \cdots \leq X_{n:n}$ of a random sample
 $X_1,\ldots,X_n$.
 They show,
 using Jacobi polynomials,
 that
 for any integers  $1\leq i < j \leq n$,
 \[
 \rho(X_{i:n},X_{j:n})\leq
 \sqrt{\frac{i(n+1-j)}{j(n+1-i)}},
 \]
 with equality if and only if the random sample arizes
 from a rectangular distribution.
 Indeed, set
 $g(u)=F^{-1}(u)=\inf\{x:F(x)\geq u\}$, $0<u<1$, where $F$ is the
 common distribution function of the i.i.d.\ r.v.'s $X_1,\ldots,X_n$,
 and consider the order statistics $U_{1:n}\leq\cdots\leq U_{n:n}$
 of a random sample $U_1,\ldots,U_n$ from ${\cal U}(0,1)$.
 Then,  $(X_{i:n},X_{j:n})\dist(g(U_{i:n}),g(U_{j:n}))$.
 Thus,
 $\rho(X_{i:n},X_{j:n})= \rho (g(U_{i:n}),g(U_{j:n}))$,
 which is well defined whenever
 $0<\Var X_{i:n} + \Var X_{j:n}<\infty$.
 Since for any $s\in(0,1)$,
 $(U_{i:n}|U_{j:n}=s) \dist \widetilde{U}_{i:j-1}$,
 where $\widetilde{U}_{i:m}$
 is the $i$-th order statistic of a sample with size
 $m$ from ${\cal U}(0,s)$, we have
 \[
 \widetilde{U}_{i:j-1}\dist s U_{i:j-1}
 \Rightarrow \E[U_{i:n}^{k}|U_{j:n}=s]
 =\E[(s U_{i:j-1})^{k}]=s^{k}\E(U_{i:j-1}^{k}).
 \]
 Now,
 \begin{eqnarray*}
 \E\big(U_{i:j-1}^{k}\big)
 & = &
 \displaystyle\int_{0}^{1}u^{k}\frac{1}{B(i,j-i)}u^{i-1}(1-u)^{j-i-1}du\\
 & = & \displaystyle\frac{B(k+i,j-i)}{B(i,j-i)}=\frac{(k+i-1)!(j-1)!}{(k+j-1)!(i-1)!}.
 \end{eqnarray*}
 In addition, for any $t \in (0,1)$ we have
 $(U_{j:n}|U_{i:n}=t)\dist \widetilde{U}_{j-i:n-i}$,
 where
 $\widetilde{U}_{j-i:n-i}$ is the $(j-i)$-th order
 statistic of a sample with size $n-i$ from
 ${\cal U}(t,1)$. Clearly,
 if $\widetilde{U}\sim{\cal U}(t,1)$
 then $\widetilde{U}\dist t+(1-t)U$ where
 $U\sim{\cal U}(0,1)$. So,
 $(U_{j:n}|U_{i:n}=t)\dist t+(1-t) U_{j-i:n-i}$
 and since $U_{j-i:n-i}\dist 1-U_{n+1-j:n-i}$, we
 get $(U_{j:n}|U_{i:n}=t)\dist 1-
 U_{n+1-j:n-i}+tU_{n+1-j:n-i}$.
 Therefore,
 \begin{eqnarray*}
 \E\big[U_{j:n}^{k}|U_{i:n}=t\big]
 \hspace{-1ex}& = &\hspace{-1ex}
 \E\big(1-U_{n+1-j:n-i}+tU_{n+1-j:n-i}\big)^{k}
 = t^{k}\E\big(U_{n+1-j:n-i}^{k}\big)+\Pol_{k-1}(t)
 \\
 \hspace{-1ex}& = &\hspace{-1ex}
 \displaystyle\frac{(n+k-j)!(n-i)!}{(n+k-i)!(n-j)!}t^{k}+\Pol_{k-1}(t).
 \end{eqnarray*}
 Thus, {\small\bf A3} is satisfied
 with
 $A_{k}
 =[i]_{k}/[j]_{k}$
 (where
 $[\alpha]_{k}:=\alpha(\alpha+1)\cdots(\alpha+k-1)$),
 and
 $B_{k}
 =[n+1-j]_{k}/[n+1-i]_{k}$.
 Clearly, $\nu=\infty$, and hence,
 \[
 \rho_{k}^{2}=A_{k}B_{k}=\frac{[i]_{k}[n+1-j]_{k}}{[j]_{k}[n+1-i]_{k}}.
 \]
 This is a strictly decreasing sequence in $k$,
 and Theorem \ref{th.1} yields the inequality
 \[
 \rho(X_{i:n},X_{j:n})\leq \sqrt{\rho_{1}^{2}}
 =\sqrt\frac{i(n+1-j)}{j(n+1-i)},
 \]
 with equality if and only if
 $g(u) \ (=F^{-1}(u))=\alpha u+\beta$ for some $\alpha>0$
 and
 $\beta\in \R$, i.e.,
 $X\sim {\cal U}(\beta, \beta+\alpha)$, $\alpha>0$.

 The same arguments apply to the case where $(X,Y)$
 has a density as in (\ref{eq.2}). Then, it is easily shown
 that for any fixed $x$ and $y$ in $(0,1)$,
 \[
 (X|Y=y)\dist y B_{\alpha,\beta} \ \ \mbox{and} \ \
 (Y|X=x)\dist x+(1-x) B_{\beta,\gamma} \dist
 1-B_{\gamma,\beta}+xB_{\gamma,\beta},
 \]
 where $B_{r,s}$ denotes a Beta r.v.\
 with parameters $r>0$ and $s>0$. It follows that
 \[
 \E(X^n|Y)=A_n Y^n \ \ \mbox{and} \ \ \E(Y^n|X)=B_n X^n +
 \Pol_{n-1}(X)
 \]
 with
 \[
 A_n=\E\big(B_{\alpha,\beta}^n\big)=\frac{[\alpha]_n}{[\alpha+\beta]_n} \ \ \mbox{and} \ \
 B_n=\E\big(B_{\gamma,\beta}^n\big)=\frac{[\gamma]_n}{[\beta+\gamma]_n}.
 \]
 Since
 $\rho_n^2=A_nB_n=\frac{[\alpha]_n[\gamma]_n}{[\alpha+\beta]_n[\beta+\gamma]_n}$
 is strictly decreasing in $n$, Theorem \ref{th.1} shows that
 $R(X,Y)= |\rho_1|=\rho_1=\rho(X,Y)$, which is identical to
 (\ref{eq.3}).
 \medskip

 \noindent
 {\small\bf Nevzorov's characterization of exponential distribution.}
 Nevzorov (1992) proved that for any $n,m\in\{1,2,\ldots\}$,
 \[
 \rho(R_{n},R_{n+m})\leq
 \sqrt{\frac{n}{n+m}},
 \]
 where $R_{i}$ is the $i$-th (upper) record
 from a continuous distribution $F$ with
 finite variance. Here $R_1=X_1$ is the first observed
 random variable
 in the i.i.d.\ sequence $\{X_i\}_{i=1}^{\infty}$. Moreover,
 equality characterizes the
 location-scale family of the standard exponential
 distribution.

 Theorem \ref{th.1} obtains Nevzorov's result immediately.
 Indeed, if $W_i$ denotes the $i$-th record from
 ${\cal E}\mbox{xp}(1)$ (with density $f(x)=e^{-x}$, $x>0$)
 then
 \[
 (W_{n},W_{n+m})\dist
 (E_{1}+\cdots+E_{n},E_{1}+\cdots+E_{n+m}),
 \ \  n,m\in\{1,2,\ldots\},
 \]
 where $\{E_{i}\}_{i=1}^{\infty}$ is an i.i.d.\ sequence from
 ${\cal E}\mbox{xp}(1)$ -- see, e.g., Arnold et al.\ (1998).
 Setting
 $X=E_{1}+\cdots+E_{n}$ and $Y=E_{1}+\cdots+E_{n+m}$,
 the joint density of
 $(X,Y)$ is
 \[
 f_{X,Y}(x,y)=\displaystyle\frac{1}{\Gamma(n)\Gamma(m)}x^{n-1}(y-x)^{m-1}e^{-y},
 \ \  0<x<y<\infty,
 \]
 and the conditional densities
 are
 \[
 f_{X|Y}(x|y)
 =\displaystyle\frac{\Gamma(n+m)}{\Gamma(n)\Gamma(m)}x^{n-1}(y-x)^{m-1}y^{-(n+m-1)},
 \ \  x\in(0,y),
 \]
 and
 \[
 f_{Y|X}(y|x)=\displaystyle\frac{1}{\Gamma(m)}(y-x)^{m-1}e^{-(y-x)},
 \ \
 y\in(x,\infty).
 \]
 It follows that
 \[
 \E(X^{k}|Y=y)
 =
 \frac{(k+n-1)!(n+m-1)!}{(k+n+m-1)!(n-1)!}y^{k}
 \]
 and
 \[
 \E(Y^{k}|X=x)=
 x^{k}+\displaystyle\frac{1}{\Gamma(m)}\sum_{i=1}^{k}\binom{k}{i}\Gamma(i+m)x^{k-i}.
 \]
 Thus, {\small\bf A3} is satisfied with
 $A_{k}=\displaystyle\frac{(k+n-1)!(n+m-1)!}{(k+n+m-1)!(n-1)!}$  and
 $B_{k}=1$, so that
 \[
 \rho_{k}^{2}=A_{k}B_{k}=\displaystyle\frac{(k+n-1)!(n+m-1)!}{(k+n+m-1)!(n-1)!}
 =\frac{[n]_k}{[n+m]_k}.
 \]
 Since this is a strictly decreasing sequence in $k$,
 Theorem \ref{th.1} yields the inequality
 \[
 \rho(R_{n},R_{n+m})=\rho
 (g(W_{n}),g(W_{n+m}))\leq \sqrt{\rho_{1}^{2}}=
 \sqrt{\frac{n}{n+m}},
 \]
 where $g(u)=F^{-1}(1-e^{-u})$, $u>0$. The equality holds
 if and only if $g$ is increasing and linear. That is,
 if and only if $F$ is the distribution function of $\alpha E+\beta$
 where $\alpha>0$, $\beta\in\R$ and
 \medskip
 $E\sim{\cal E}\mbox{xp}(1)$.

 \noindent
 {\small\bf L\'{o}pez-Bl\'{a}zquez and Casta\~{n}o-Mart\'{i}nez result
 on maximal correlation of order statistics from a finite population.}
 Let
 $U_{1:n}^{(N)}<U_{2:n}^{(N)} <\cdots<U_{n:n}^{(N)}$
 be the order statistics corresponding to a simple random
 sample,
 $U_1^{(N)},\ldots,U_n^{(N)}$, taken without replacement from the
 finite ordered population $\Pi_N=\{1,2,\ldots,N\}$, where $2\leq
 n<N$. Since
 $\Pr(U_{i:n}^{(N)}=k)={k-1\choose i-1}{N-k\choose n-i}{N\choose n}^{-1}$
 for $k\in\{i,i+1,\ldots,N-(n-i)\}$ (and $0$ otherwise), and this defines a probability
 mass function with support $A_{i:n}^{(N)}:=\{i,i+1,\ldots,N-(n-i)\}$, we
 conclude the identity
 \be
 \label{eq.14}
 \sum_{k=i}^{N-(n-i)}{k-1\choose i-1}{N-k\choose n-i}
 ={N\choose n}, \ \ 1\leq i\leq n\leq N.
 \ee
 Setting $[\alpha]_m=\alpha (\alpha+1)\cdots (\alpha+m-1)$ (with
 $[\alpha]_0=1$ for all $\alpha\in\R$) we can derive, with the help
 of (\ref{eq.14}), a simple expression for
 the ascending moments of $U_{i:n}^{(N)}$:
 \be
 \label{eq.15}
 \E\big\{[U_{i:n}^{(N)}]_m\big\}=[N+1]_m \frac{[i]_m}{[n+1]_m}, \ \
 m=1,2,\ldots \ .
 \ee
 We also mention the following obvious relations, holding for all
 $1\leq i<j\leq n$:
 \begin{eqnarray}
 \label{eq.16}
 (U_{i:n}^{(N)}, U_{j:n}^{(N)})
 \hspace{-1ex}& \dist &\hspace{-1ex}
 (N+1-U_{n+1-i:n}^{(N)}, N+1-U_{n+1-j:n}^{(N)}),
 \\
 \label{eq.17}
 (U_{i:n}^{(N)} \big| U_{j:n}^{(N)}=s)
 \hspace{-1ex}& \dist &\hspace{-1ex}
 U_{i:j-1}^{(s-1)}, \ \ s\in\{j,j+1,\ldots,N-(n-j)\},
 \\
 \label{eq.18}
 (U_{j:n}^{(N)} \big| U_{i:n}^{(N)}=k)
 \hspace{-1ex}& \dist &\hspace{-1ex}
 k+U_{j-i:n-i}^{(N-k)}, \ \ k\in\{i,i+1,\ldots,N-(n-i)\}.
 \end{eqnarray}
 Now, applying (\ref{eq.15}) and (\ref{eq.17}) we have
 \be
 \label{eq.19}
 \E\big\{ [U_{i:n}^{(N)}]_{m} \big| U_{j:n}^{(N)}=s \big\}
 =[s]_m \frac{[i]_m}{[j]_m}, \ \ m=1,2,\ldots \ .
 \ee
 Let $(X,Y)=(U_{i:n}^{(N)}, U_{j:n}^{(N)})$
 and observe that $\nu_X=\nu_Y=N-n\geq 1$; see Remark \ref{rem.new}.
 Relation (\ref{eq.19})
 shows that
 \[
 \E\big([X]_m|Y\big)=\frac{[i]_m}{[j]_m}[Y]_m=\frac{[i]_m}{[j]_m}Y^m
 +\Pol_{m-1}(Y), \ \ m=1,2,\ldots,
 \]
 which implies, using induction on $m$, that
 \be
 \label{eq.20}
 \E(X^m|Y)=\frac{[i]_m}{[j]_m}Y^m
 +\Pol_{m-1}(Y), \ \ m=1,2,\ldots \ .
 \ee
 Similarly, set $i'=n+1-j$, $j'=n+1-i$
 (so that $1\leq i'<j'\leq n$) and write
 $U_{i'}$ instead of $U_{i':n}^{(N)}$,
 $U_{j'}$ instead of $U_{j':n}^{(N)}$.
 Now, applying relations (\ref{eq.16}) and (\ref{eq.19}),
 \begin{eqnarray*}
 \E\big([Y]_m|X=k\big)
 \hspace{-1ex}&=&\hspace{-1ex}
 \E\left\{[N+1-U_{i'}]_m|U_{j'}=N+1-k\right\}
 \\
 \hspace{-1ex}&=&\hspace{-1ex}
 \E\left\{(-1)^m [U_{i'}]_m+\Pol_{m-1}(U_{i'}) \ | \ U_{j'}=N+1-k\right\}
 \\
 \hspace{-1ex}&=&\hspace{-1ex}
 (-1)^m\E\left\{[U_{i'}]_m \ | \ U_{j'}=N+1-k\right\}
 +\Pol_{m-1}(N+1-k)
 \\
 \hspace{-1ex}&=&\hspace{-1ex}
 (-1)^m [N+1-k]_m \frac{[i']_m}{[j']_m} +\Pol_{m-1}(k)
 \\
 \hspace{-1ex}&=&\hspace{-1ex}
 [k]_m \frac{[i']_m}{[j']_m} +\Pol_{m-1}(k)
 =[k]_m \frac{[n+1-j]_m}{[n+1-i]_m} +\Pol_{m-1}(k).
 \end{eqnarray*}
 It follows that
 $\E([Y]_m|X)=\frac{[n+1-j]_m}{[n+1-i]_m} [X]_m +\Pol_{m-1}(X)
 =\frac{[n+1-j]_m}{[n+1-i]_m} X^m +\Pol_{m-1}(X)$
 and,  using induction on $m$,
 \be
 \label{eq.21}
 \E(Y^m|X)
 =\frac{[n+1-j]_m}{[n+1-i]_m} X^m +\Pol_{m-1}(X),
 \ \ m=1,2,\ldots \ .
 \ee
 Clearly, (\ref{eq.20}) and (\ref{eq.21}) show that
 {\small\bf A3} is satisfied for $(X,Y)$. Moreover, we have found that
 $A_m=[i]_m/[j]_m$ and
 $B_m=[n+1-j]_m/[n+1-i]_m$,
 both of which do not depend on $N$.
 Since $A_m>0$ and $\rho_m=\sqrt{A_m B_m{\bf1}_{\{m\leq N-n\}}}$
 is strictly decreasing in $m\in\{1,\ldots,N-n,N-n+1\}$,
 Theorem \ref{th.1} yields the inequality
 \[
 \rho\big(g_1(U_{i:n}^{(N)}),g_2(U_{j:n}^{(N)})\big) \leq
 \sqrt{\rho_1^2}=\sqrt{\frac{i(n+1-j)}{j(n+1-i)}}.
 \]
 The equality holds if and only if both $g_1$ and $g_2$ are
 (non-constant and)
 linear and with the same monotonicity. More precisely,
 the restriction of $g_1$ in the set
 $A_{i:n}^{(N)}$
 has to be non-constant and linear and the restriction
 of $g_2$ in the set  $A_{j:n}^{(N)}$ has to be non-constant, linear
 and with the same monotonicity as $g_1$. Note that
 both sets $A_{i:n}^{(N)}$ and $A_{j:n}^{(N)}$ contain at least two points
 if and only if $N\geq n+1$.

 Lemma 2.1 of Balakrishnan et al.\
 (2003) asserts that for the non-decreasing function
 $g:\{1,2,\ldots,N\}\to\{x_1\leq x_2\leq \cdots\leq x_N\}:=\widetilde{\Pi}_N$
 with $g(i)=x_i$, $i=1,2,\ldots,N$,
 \[
 \big(g(U_{i:n}^{(N)}),g(U_{j:n}^{(N)})\big)\dist (X_{i:n},
 X_{j:n}), \ \ 1\leq i<j\leq n,
 \]
 where
 $X_{1:n}\leq X_{2:n}\leq \cdots \leq X_{n:n}$
 are the order statistics corresponding to a simple random sample
 drawn (without replacement) from the finite population
 $\widetilde{\Pi}_N$.  Suppose that
 $\rho(X_{i:n},X_{j:n})$ is well-defined or, equivalently,
 that the elements of $\widetilde{\Pi}_N$ satisfy
 $x_i<x_{N-(n-i)}$ and $x_j<x_{N-(n-j)}$
 (otherwise, at least one of $X_{i:n}, X_{j:n}$ would be degenerate).
 Then we conclude that
 \be
 \label{eq.22}
 \rho(X_{i:n},X_{j:n})\leq  \sqrt{\frac{i(n+1-j)}{j(n+1-i)}},
 \ \
 1\leq i<j\leq n<N.
 \ee
 The equality,
 for fixed $i$, $j$, $n$, $N$,
 characterizes those finite populations
 $\widetilde{\Pi}_N$ for which the sets
 $\{x_i,x_{i+1},\ldots,x_{N-(n-i)}\}$ and
 $\{x_j,x_{j+1},\ldots,x_{N-(n-j)}\}$,
 which may or may not have common points,
 consist of consecutive terms of two (possibly different)
 strictly increasing arithmetic progresses.
 That is, a population of size $N$ with elements
 $x_1\leq x_2\leq \cdots \leq x_N$
 satisfying $x_i<x_{N-(n-i)}$ and
 $x_j<x_{N-(n-j)}$ attaints the equality in (\ref{eq.22})
 if and only if there exist constants $a_1>0$, $b_1\in\R$, $a_2>0$ and $b_2\in \R$
 such that
 \[
 x_k=\left\{
 \begin{array}{cl}
 a_1 k+b_1, & \mbox{for } k=i,i+1,\ldots,N-(n-i),
 \\
 a_2 k+b_2, & \mbox{for } k=j,j+1,\ldots,N-(n-j),
 \\
 \mbox{arbitrary}, & \mbox{otherwise.}
 \end{array}
 \right.
 \]
 L\'{o}pez-Bl\'{a}zquez and
 Casta\~{n}o-Mart\'{i}nez (2006), using Hahn polynomials,
 have obtained a corresponding inequality for the correlation
 ratio, which implies inequality (\ref{eq.22}). Their arguments,
 however, apply to populations $\widetilde{\Pi}_N$ having
 $N$ distinct elements. We also refer to Theorem 2.1 and Corollary 2.1 in
 Casta\~{n}o-Mart\'{i}nez et al.\ (2007), noting
 that the characterization result stated in Corollary 2.1
 of this article is incomplete, unless the sets $A_{i:n}^{(N)}$ and
 $A_{j:n}^{(N)}$ have at least two common points, i.e.,
 $N\geq n+(j-i)+1$.

 \section{Records from a splitting model and a
 Nevzorov-type characterization of the exponential
 distribution}
 \label{sec.last}
 Assume that in a particular country and for a specific athletic event,
 the consecutive performances of the athletes are described by an i.i.d.\ sequence
 $\{X_i\}_{i=1}^{\infty}$. Here and elsewhere in this section,
 the common distribution of
 each $X_i$ will be assumed to be continuous, i.e., with no
 atoms -- absolute continuity is not required.
 As the time goes on, the common practice is that some data regarding
 the sequence of national records, i.e., the sequence $\{R_i\}_{i=1}^{\infty}$,
 are recorded, in contrast to the
 original performances of the athletes,
 $X_i$, which are usually lost or forgotten.
 The above considerations  give rise to the classical record
 model, based on an i.i.d.\ sequence, which is well-developed in the
 literature; see Arnold et al.\ (1998).
 Under this classical model the
 observed sequence $\{R_i\}_{i=1}^n$ of the first
 $n$ upper national records is defined as $R_1=X_1$ and
 $R_{i}=X_{T(i)}$, $i=2,\ldots,n$, where
 $T(i)=\min\{j\in\{1,2,\ldots\}:X_j>R_{i-1}\}$.

 Suppose now that, after the appearance of the $n$-th national record,
 the initial country is divided into, say, two new countries (branches),
 and assume that the athletes in each country are of the same
 strength as they were before the division. Then, the subsequent
 national records
 in each branch will take into account the current (common) national record, $R_n$,
 and the subsequent sequence of their individual records will
 be of the form $(R'_{n+n_1}, R''_{n+n_2})$, with
 $n_1,n_2\in\{1,2,\ldots\}$. Clearly,
 \be
 \label{eq.records}
 R'_{n+n_1}\dist R_{n+n_1} \ \ \mbox{and} \ \ R''_{n+n_2}\dist R_{n+n_2}
 \ee
 where $R_{n+m}$ is the $(n+m)$-th record from the initial
 sequence, but as $n_1$ and $n_2$ become large, the r.v.'s
 $R'_{n+n_1}$ and $R''_{n+n_2}$ should tend towards
 independence.

 Thus, the actual definition of the splitting record sequence
 is equivalent
 to the following model: Let
 $\{X_1,X'_1,X''_1,X_2,X_2',X_2'',\ldots\}$ be an i.i.d.\
 sequence of r.v.'s.
 Define the $n$-th upper record $R_n$ as
 before (based on the $X_i$'s), then set $R_n'=R_n'':=R_n$ and
 $T'(n)=T''(n):=T(n)$. For $i=1,2,\ldots$
 define the subsequent record times and record values
 by
 \begin{eqnarray*}
 && T'(n+i)=\min\{j\in\{1,2,\ldots\}:X'_j>R'_{n+i-1}\},
 \ R'_{n+i}=X'_{T'(n+i)}, \ \ \mbox{and}
 \\
 && T''(n+i)=\min\{j\in\{1,2,\ldots\}:X''_j>R''_{n+i-1}\},
 \ R''_{n+i}=X''_{T''(n+i)}.
 \end{eqnarray*}
 Clearly, it is of some interest to study
 the correlation behavior of the marginal records under
 this model, since large correlation among these variables
 entails good prediction of one branch to the other.
 It is not surprising that, similarly to the classic case,
 the splitting record sequence
 satisfies several interesting properties.
 In particular, in what follows we shall make use
 of the following
 lemma.
 \begin{LEM}
 \label{lem.3}
 (a) Let $\{(W'_{n+n_1},W''_{n+n_2})\}_{n_1,n_2=1}^{\infty}$
 be the splitting record sequence
 based on the i.i.d.\ sequence $\{E_i,E'_i,E''_i\}_{i=1}^{\infty}$
 from the standard exponential
 distribution, ${\cal E}\mbox{xp}(1)$.
 Then
 for each $n_1,n_2\in\{1,2,\ldots\}$,
 \be
 \label{eq.24}
 (W'_{n+n_1},W''_{n+n_2})\dist(E_1+\cdots+E_n+E'_{1}+\cdots +E'_{n_1},
 E_1+\cdots+E_n+E''_{1}+\cdots +E''_{n_2}).
 \ee
 (b)
 Let $\{(R'_{n+n_1},R''_{n+n_2})\}_{n_1,n_2=1}^{\infty}$ be the splitting record sequence
 based on the i.i.d.\ sequence $\{X_i,X'_i,X''_i\}_{i=1}^{\infty}$
 from a non-atomic (continuous) distribution function $F$.
 Then, for each $n_1,n_2\in\{1,2,\ldots\}$,
 \be
 \label{eq.25}
 (R'_{n+n_1},R''_{n+n_2})\dist (g(W'_{n+n_1}),g(W''_{n+n_2})),
 \ee
 where $g(u)=F^{-1}(1-e^{-u})$, $u>0$, with $F^{-1}(y)
 =\inf\{x:F(x)\geq y\}$, $y\in(0,1)$.
 \end{LEM}

 The proof of Lemma \ref{lem.3} is simple and is
 left to the reader -- cf.\ Arnold et al.\ (1998).
 With the help of this lemma, Theorem
 \ref{th.1} yields the following characterization.
 \begin{THEO}
 \label{th.2}
 If $(R'_{n+n_1},R''_{n+n_2})$ are splitting records based on an
 i.i.d.\ sequence $\{X_i\}_{i=1}^{\infty}$ from a non-atomic
 distribution $F$ with
 $\E (R'_{n+n_1})^2<\infty$ and $\E (R''_{n+n_2})^2<\infty$
 then
 \[
 \rho(R'_{n+n_1},R''_{n+n_2})\leq
 \frac{n}{\sqrt{n+n_1}\sqrt{n+n_2}}.
 \]
 The equality holds if and only if $F$ is the distribution
 function of $\alpha E+\beta$ for some $\alpha>0$ and $\beta\in\R$,
 where $E\sim{\cal E}\mbox{xp}(1)$.
 \end{THEO}

 \noindent
 \begin{Proof}
 Set $X=E_1+\cdots+E_n+E'_1+\cdots+E'_{n_1}$ and
 $Y=E_1+\cdots+E_n+E''_1+\cdots+E''_{n_2}$
 with $(E_1,\ldots,E''_{n_2})$  being a vector of
 $n+n_1+n_2$ i.i.d.\ standard exponential r.v.'s.
 It can be shown (see the proof of Theorem \ref{th.3}, below)
 that for all
 $k\in\{1,2,\ldots\}$,
 \[
 \E(X^k|Y)=\frac{[n]_k}{[n+n_2]_k}Y^k+\Pol_{k-1}(Y),
 \ \ \
 \E(Y^k|X)=\frac{[n]_k}{[n+n_1]_k}X^k+\Pol_{k-1}(X).
 \]
 That is, the random vector $(X,Y)$
 has the polynomial regression property
 with $A_k=[n]_k/[n+n_2]_k$ and
 $B_k=[n]_k/[n+n_1]_k$. Clearly,
 $\rho_k^2=([n]_k)^2/([n+n_1]_k[n+n_2]_k)$ is
 strictly decreasing in $k$.
 In view of Lemma \ref{lem.3}, Theorem \ref{th.1}
 shows that, with $g(u)=F^{-1}(1-e^{-u})$,
 \begin{eqnarray*}
 \rho(R'_{n+n_1},R''_{n+n_2})
 \hspace{-1ex}&=&\hspace{-1ex}
 \rho\big(g(W'_{n+n_1}),g(W''_{n+n_2})\big)
 \\
 \hspace{-1ex}&=&\hspace{-1ex}
 \rho(g(X),g(Y))\leq
 \sqrt{\rho_1^2}=\frac{n}{\sqrt{n+n_1}\sqrt{n+n_2}},
 \end{eqnarray*}
 where the equality holds if and only if $g:(0,\infty)\to\R$ is
 linear.
 This, together with the fact that $g$ has
 assumed to be strictly increasing,
 completes the proof.
 \medskip
 $\Box$
 \end{Proof}

 Provided that every component
 is representative as a sum on independent gamma r.v.'s
 with the same scale parameter, say $1/\lambda$,
 Theorem \ref{th.2} and Nevzorov's (1992)
 characterization reflect the polynomial regression
 property of a specific class of multivariate gamma
 random vectors.
 Recall that a random variable $X$ follows a gamma
 distribution with parameters $\alpha>0$ and $\lambda>0$
 if its density is given by
 \[
 f(x)=\frac{\lambda^\alpha}{\Gamma(\alpha)}x^{\alpha-1}e^{-\lambda x}, \ \
 x>0.
 \]
 This is denoted by $X\sim\Gamma(\alpha;\lambda)$, while
 the notation $X\sim \Gamma(0;\lambda)$ (for some $\lambda>0$)
 means that $X$ is degenerate and takes the value
 zero w.p.\ $1$. In any case, $\E X =\alpha/\lambda$ and
 $\Var X=\alpha/\lambda^2$.
 Under the above notation one can easily verify the following
 result,
 which
 contains both Theorem \ref{th.2} and
 Nevzorov's characterization as particular cases.
 In fact, Theorem \ref{th.3}
 obtains
 the maximal correlation of
 Cheriyan's bivariate gamma distribution --
 see Cheriyan (1941) and Balakrishnan and Lai (2009), pp.\
 322--325.
 \begin{THEO}
 \label{th.3}
 Let $X_i\sim\Gamma(\alpha_i;\lambda)$ ($i=0,1,2$)
 be independent r.v.'s
 with
 $\lambda>0$,
 $\alpha_i\geq 0$ ($i=0,1,2$)
 and
 $\alpha_0+\alpha_i>0$ ($i=1,2$).
 Then the random vector $(X,Y)=(X_0+X_1,X_0+X_2)$
 follows a bivariate distribution with gamma marginals,
 namely $X\sim \Gamma(\alpha_0+\alpha_1;\lambda)$
 and $Y\sim \Gamma(\alpha_0+\alpha_2;\lambda)$.
 Moreover, $(X,Y)$ satisfies the polynomial regression
 property. More precisely, for all $n\in\{1,2,\ldots\}$,
 \[
 \E(X^n|Y)=\sum_{j=0}^{n}{n\choose j}
 \frac{[\alpha_0]_j[\alpha_1]_{n-j}}{\lambda^{n-j}[\alpha_0+\alpha_2]_j}Y^j,
 \ \
 \E(Y^n|X)=\sum_{j=0}^{n}{n\choose j}
 \frac{[\alpha_0]_j[\alpha_2]_{n-j}}{\lambda^{n-j}[\alpha_0+\alpha_1]_j}X^j,
 \]
 where $[\alpha]_0\equiv1$ for all $\alpha\in\R$
 and $[\alpha]_k=\alpha(\alpha+1)\cdots(\alpha+k-1)$
 for $k\in\{1,2,\ldots\}$. Finally,
 for any $g_1\in L^2(X)$ with $\Var g_1(X)>0$
 and for any $g_2\in L^2(Y)$ with $\Var g_2(Y)>0$
 we have the inequality
 \[
 \rho(g_1(X),g_2(Y))\leq
 \frac{\alpha_0}{\sqrt{\alpha_0+\alpha_1}\sqrt{\alpha_0+\alpha_2}}.
 \]
 Provided that $\alpha_1+\alpha_2>0$, the equality holds if
 and only if either (i) $\alpha_0=0$
 and $g_1$, $g_2$ are arbitrary or (ii)
 $\alpha_0>0$ and both $g_1$,  $g_2$ are nonconstant, linear
 and with the same monotonicity.
 \end{THEO}

 \noindent
 \begin{Proof}
 Cases $\alpha_0=0$ and $\alpha_1=\alpha_2=0$ are simple ($X$,$Y$ are
 independent and $X=Y$ w.p.\ 1, respectively).
 Both cases $\alpha_0>0$, $\alpha_1=0$, $\alpha_2>0$
 and $\alpha_0>0$, $\alpha_1>0$, $\alpha_2=0$ are similar
 to Nevzorov's case and can be shown as in Section
 \ref{sec.known}. Assume now that $\alpha_i>0$
 for $i=0,1,2$. Then, it is easily shown that the conditional
 density of $X$ given $Y=y$ (for any fixed $y>0$) is
 \hspace{-.7ex}
 \[
 f_{X|Y}(x|y)=c
 e^{-\lambda x}
 \int_{0}^{\min\{x,y\}}
 w^{\alpha_0-1}
 (x-w)^{\alpha_1-1}(y-w)^{\alpha_2-1}e^{\lambda w}
 dw,
 \ \ \ x>0,
 \hspace{-.7ex}
 \]
 where
 \[
 c=c(\alpha_0,\alpha_1,\alpha_2;\lambda;y)=\frac{\lambda^{\alpha_1}
 \Gamma(\alpha_0+\alpha_2)}{y^{\alpha_0+\alpha_2-1}
 \Gamma(\alpha_0)\Gamma(\alpha_1)\Gamma(\alpha_2)}.
 \]
 Despite the fact that this conditional density is not given in a closed
 form, we can calculate $\E(X^n|Y=y)$ using Tonelli's Theorem. Indeed,
 consider the nonnegative functions $\theta(w)=w^{\alpha_0-1}e^{\lambda w}$ ($w>0$) and
 $h(x,y,w)=(x-w)^{\alpha_1-1}(y-w)^{\alpha_2-1}{\bf1}_{\{w<\min\{x,y\}\}}$ ($x,y,w>0$).
 Then,
 \hspace{-.7ex}
 \begin{eqnarray*}
 \E(X^n|Y=y)
 \hspace{-1ex}&=&\hspace{-1ex}
 c\Big\{ \int_0^y x^n e^{-\lambda x} \int_{0}^x \theta(w) h(x,y,w)dw dx
 \Big.
 \hspace{-.7ex}
 \\
 &&
 \hspace{15ex}+
 \Big.
 \int_y^{\infty} x^n e^{-\lambda x} \int_{0}^y \theta(w) h(x,y,w)dw dx \Big\}
 \\
 \hspace{-1ex}&=&\hspace{-1ex}
 c\Big\{
 \int_0^y \theta(w) \int_{w}^y x^n e^{-\lambda x} h(x,y,w)dx dw
 \Big.
 \hspace{-.7ex}
 \\
 &&
 \hspace{15ex}+
 \Big.
 \int_0^{y} \theta(w) \int_{y}^{\infty} x^n e^{-\lambda x}
 h(x,y,w)dx
 dw \Big\}
 \\
 \hspace{-1ex}&=&\hspace{-1ex}
 c\int_0^{y} \theta(w) \int_{w}^{\infty} x^n e^{-\lambda x}
 h(x,y,w)dx dw
 \\
 \hspace{-1ex}&=&\hspace{-1ex}
 c\int_0^{y} w^{\alpha_0-1}
 (y-w)^{\alpha_2-1} \Big\{\int_{0}^{\infty} (x+w)^n e^{-\lambda x}
 x^{\alpha_1-1}dx\Big\} dw.
 \end{eqnarray*}
 Now, expanding $(x+w)^n$ according to Newton's formula
 and using
 $\int_{0}^{\infty}x^{j+\alpha_1-1}e^{-\lambda x}dx
 =\Gamma(\alpha_1+j)/\lambda^{\alpha_1+j}$
 ($j=0,1,\ldots,n$),
  the inner integral is equal to
 \[
 \int_{0}^{\infty} (x+w)^n e^{-\lambda x}
 x^{\alpha_1-1}dx
 =\frac{1}{\lambda^{\alpha_1}}\sum_{j=0}^n {n\choose j}
 \frac{\Gamma(\alpha_1+j)}{\lambda^{j}} w^{n-j}.
 \]
 Substituting this expression to the double integral,
 we obtain
 \begin{eqnarray*}
 \E(X^n|Y=y)
 \hspace{-1ex}&=&\hspace{-1ex}
 \frac{c}{\lambda^{\alpha_1}}\sum_{j=0}^n {n\choose j}
 \frac{\Gamma(\alpha_1+j)}{\lambda^{j}}
 \int_0^{y} w^{\alpha_0+(n-j)-1}
 (y-w)^{\alpha_2-1} dw
 \\
 \hspace{-1ex}&=&\hspace{-1ex}
 \frac{c}{\lambda^{\alpha_1}}\sum_{j=0}^n {n\choose j}
 \frac{\Gamma(\alpha_1+j)}{\lambda^{j}}
 \frac{\Gamma(\alpha_2)\Gamma(\alpha_0+n-j)}{\Gamma(\alpha_0+\alpha_2+n-j)}
 y^{\alpha_0+\alpha_2+(n-j)-1}
 \\
 \hspace{-1ex}&=&\hspace{-1ex}
 \frac{\Gamma(\alpha_0+\alpha_2)}{\Gamma(\alpha_0)\Gamma(\alpha_1)}
 \sum_{j=0}^n {n\choose j}
 \frac{\Gamma(\alpha_0+j)\Gamma(\alpha_1+n-j)}
 {\lambda^{n-j}\Gamma(\alpha_0+\alpha_2+j)}
 y^{j}.
 \end{eqnarray*}
 Therefore, $X$ has polynomial regression on $Y$ and, similarly,
 $Y$ has polynomial regression on $X$. It follows that
 $(X,Y)$ satisfies conditions {\small\bf A1}--{\small\bf A3} and,
 \vspace{-.7ex}
 moreover,
  \[
 \rho_n=\sign(A_n)\sqrt{A_nB_n}=\frac{[\alpha_0]_n}
 {\sqrt{[\alpha_0+\alpha_1]_n}\sqrt{[\alpha_0+\alpha_2]_n}}.
 \vspace{-.7ex}
 \]
 Since $|\rho_n|=\rho_n$ is strictly decreasing in $n$, a
 final application of Theorem \ref{th.1}
 completes the proof.
 \medskip
 $\Box$
 \end{Proof}

 Theorem \ref{th.3} includes
 Nevzorov's (1992) characterization because, taking
 $\lambda=1$, $\alpha_0=n$, $\alpha_1=0$, $\alpha_2=m$
 and $g_1(u)=g_2(u)=F^{-1}(1-e^{-u})$, $u>0$,
 we have that, under the standard record
 model,
 $
 (R_n,R_{n+m})\dist(g(W_n),g(W_{n+m}))\dist(g(X),g(Y)).
 $
 Here $(W_n,W_{n+m})$ are the corresponding upper records from
 the standard exponential distribution. Clearly, the theorem also
 includes the result on splitting record models of Theorem
 \ref{th.2} -- the only difference being that, due to Lemma \ref{lem.3},
 one has now to put $\alpha_1=n_1$ (rather than $\alpha_1=0$)
 and $\alpha_2=n_2$ (rather than $\alpha_2=m$).

 Provided that $g_1,g_2\in C^{\infty}(0,\infty)$,
 $g_1(X)\in L^2(X)$, $g_2(Y)\in L^2(Y)$, and
 assuming that $\E\big|X^n g_1^{(n)}(X)\big|<\infty$ and
 $\E\big|Y^n g_2^{(n)}(Y)\big|<\infty$ for all $n$,
 where $g_i^{(n)}$ denotes the $n$-th derivative of
 $g_i$, $i=1,2$,
 it is of some interest to note that (\ref{eq.10})
 yields the covariance identity
 \vspace{-.7ex}
 \be
 \label{eq.26}
 \Cov[g_1(X),g_2(Y)]=\sum_{n=1}^{\infty}
 \frac{[\alpha_0]_n}{n![\alpha_0+\alpha_1]_n
 [\alpha_0+\alpha_2]_n} \E\big(X^n g_1^{(n)}(X)\big)
 \E\big(Y^n g_2^{(n)}(Y)\big).
 \vspace{-.7ex}
 \ee
 Of course one can apply (\ref{eq.26})
 to the case $\alpha_1=\alpha_2=0$, $\alpha_0>0$.
 Then, $X=Y\sim \Gamma(\alpha_0;\lambda)$
 and we reobtain the generalized Stein-type identity for
 the $\Gamma(\alpha_0;\lambda)$
 \vspace{-.7ex}
 distribution (see Afendras et al.\
 (2011)):
 \be
 \label{eq.27}
 \Cov[g_1(X),g_2(X)]=\sum_{n=1}^{\infty}
 \frac{1}{n![\alpha_0]_n}
 \E\big(X^n g_1^{(n)}(X)\big) \E\big(X^n g_2^{(n)}(X)\big).
 \vspace{-.7ex}
 \ee
 Similarly, we can apply (\ref{eq.26}) to the classical record setup
 from the standard exponential (setting $\lambda=1$, $\alpha_0=n$,
 $\alpha_1=0$ and $\alpha_2=m$). Then we get
 \vspace{-.7ex}
 \be
 \label{eq.28}
 \hspace{-1.5ex}
 \Cov[g_1(W_n),g_2(W_{n+m})]=\sum_{k=1}^{\infty}
 \frac{1}{k![n+m]_k}
 \E\big(W_n^k g_1^{(k)}(W_n)\big) \E\big(W_{n+m}^k g_2^{(k)}(W_{n+m})\big).
 \vspace{-.7ex}
 \ee

 \section{Conclusions}
 \label{sec.last2}
 The simplicity of the proposed method depends
 heavily on the polynomial regression property,
 {\small\bf A3}, which is
 satisfied by all bivariate distributions
 discussed
 in the present article.
 Incidentally, in all of our cases we
 concluded that $R=|\rho(X,Y)|=\sqrt{A_1B_1}$,
 and some times it is asserted that this is the typical
 situation whenever ${\small\bf A3}$
 is merely satisfied for $n=1$ (i.e.,
 when both variables have linear regression).
 However, this is not true, e.g., when
 $(X,Y)$ is uniformly distributed on
 the interior of the unit disc (then $A_1=B_1=\rho(X,Y)=0$);
 see, also, Dembo et al.\ (2001).

 Casta\~{n}o-Mart\'{i}nez et al.\ (2007)
 develop a correlation model for partial minima
 (or maxima) rather than records. Their Section
 3 indicates that many difficulties can enter to the correlation
 problem when {\small\bf A3} fails. It appears that, in such cases, one
 has to calculate the values of $\rho_{n,k}: =\E[\phi_n(X)\psi_k(Y)]$
 for all $n$ and $k$. This is not an easy task in general, in contrast to
 the present simplified situation,
 where knowledge of the values $A_n$ and $B_n$
 in {\small\bf A3} suffices for the calculation of the maximal
 correlation coefficient.
 \medskip

 \noindent
 {\small\bf Acknowledgement.} We are grateful to an
 anonymous referee for pointing out a serious error
 in the previous version of our main result, Theorem \ref{th.1},
 regarding the finite support case.

 \vspace{1em}

 \begin{center}
 {{\bf REFERENCES} }
 \end{center}
 \vspace{-1em}

 {\small

 \begin{description}

 \item Afendras, G.; Papadatos, N.; Papathanasiou, V.\
 (2011). An extended Stein-type covariance identity for
 the Pearson family, with applications to lower variance bounds,
 {\it Bernoulli}, {\bf 17}, 507--529.
 \vspace{-1.0ex}

 \item Arnold, B.C.; Balakrishnan, N.; Nagaraja, H.N.\
 (1998). {\it Records}. Wiley, New York.
 \vspace{-1.0ex}

 \item Balakrishnan, N.; Charalambides, C.; Papadatos, N.\
 (2003). Bounds on expectation of order statistics from a finite
 population,
 {\it J.\ Statist.\ Plann.\ Inference}, {\bf 113}, 569--588.
 \vspace{-1.0ex}

 \item Balakrishnan, N.; Lai, C.D.\
 (2009). {\it Continuous Bivariate Distributions}. Springer,
 New York (2nd ed.).
 \vspace{-1.0ex}

 \item Breiman, L.; Friedman, J.H.\ (1985). Estimating
 optimal transformations for multiple regression and
 correlation, {\it J.\ Amer.\ Statist.\ Assoc.},
 {\bf 80}, 590--598.
 \vspace{-1.0ex}

 \item Bryc, W.; Dembo, A.; Kagan, A.\
 (2005). On the maximum correlation coefficient,
 {\it Theory Probab.\ Appl.}, {\bf 49}, 132--138.
 \vspace{-1.0ex}

 \item Casta\~{n}o-Mart\'{i}nez, A.; L\'{o}pez-Bl\'{a}zquez, F.;
 Salamanca-Mi\~{n}o, B.\ (2007).
 Maximal correlation between order statistics.
 In: {\it Recent Developments in Ordered Random Variables},
 M.\ Ahsanullah and M.\ Raqab (eds.), Nova Science
 Publishers, 55--68.
 \vspace{-1.0ex}

 \item Cheriyan, K.C.\ (1941). A bivariate correlated gamma-type
 distribution function.
 {\it J.\ Indian Math.\ Soc.}, {\bf 5}, 133--144.
 \vspace{-1.0ex}

 \item Dembo, A.; Kagan, A.; Shepp., L.A.\
 (2001). Remarks on the maximum correlation coefficient,
 {\it Bernoulli}, {\bf 7}, 343--350.
 \vspace{-1.0ex}

 \item Gebelein, H.\ (1941). Das Statistische Problem der
 Korrelation als Variation
 und Eigenwertproblem und sein Zusammenhang mit
 der Ausgleichrechnung, {\it Angew.\ Math.\ Mech.}, {\bf 21}, 364--379.
 \vspace{-1.0ex}

 \item Koudou, A.E.\ (1996). Probabilit\'{e}s de Lancaster.
 {\it Exp.\ Math.}, {\bf 14}, 247--275.
 \vspace{-1.0ex}

 \item Koudou, A.E.\ (1998). Lancaster bivariate probability distributions
 with Poisson, negative binomial and gamma margins.
 {\it Test}, {\bf 7}, 95--110.
 \vspace{-1.0ex}

 \item Lancaster, H.O.\ (1957). Some properties of the
 bivariate normal distribution considered in the form of
 a contingency table, {\it Biometrika}, {\bf 44}, 289--292.
 \vspace{-1.0ex}

 \item Lancaster, H.O.\ (1969).
 {\it The chi-squared distribution}, Wiley, New York.
 \vspace{-1.0ex}

 \item Liu, J.S.; Wong, W.H.; Kong, A.\ (1994). Covariance
 structure of the Gibbs sampler with applications to the
 comparisons of estimators and augmentation schemes,
 {\it Biometrika}, {\bf 81}, 27--40.
 \vspace{-1.0ex}

 \item L\'{o}pez-Bl\'{a}zquez, F.; Casta\~{n}o-Mart\'{i}nez, A.\
 (2006). Upper and lower bounds for the correlation ratio of order
 statistics from a sample without replacement, {\it J.\
 Statist.\ Plann.\ Inference}, {\bf 136}, 43--52.
 \vspace{-1.0ex}

 \item Nevzorov, V.B.\ (1992). A characterization of exponential distributions
 by correlations between records,
 {\it  Math.\ Meth.\ Statist.}, {\bf 1}, 49--54.
 \vspace{-1.0ex}

 \item Sz\'{e}kely, G.J.; M\'{o}ri, T.F.\ (1985).
 An extremal property of rectangular
 distributions, {\it Statist.\ Probab.\ Lett.}, {\bf 3},
 107--109.
 \vspace{-1.0ex}

 \item Terrell, G.R.\ (1983). A characterization of rectangular
 distributions,
 {\it Ann.\ Probab.}, {\bf 11}, 823--826.
 \vspace{-1.0ex}

 \item Yu, Y.\ (2008). On the maximal correlation
 coefficient,
 {\it Statist.\ Probab.\ Lett.}, {\bf 78}, 1072--1075.

 \end{description}
 }

 \end{document}